# Quantifying Social Presence in Mixed Reality: A Contemporary Review of Techniques and Innovations

**Sparsh Srivastava**
ss6381@columbia.edu

**Department of Computer Science**
*Columbia University in the City of New York*
New York City, NY, 10025

**ABSTRACT**

This literature review investigates the transformative potential of mixed reality (MR) technology, where we explore the intersection of contemporary technological advancements, modern deep learning recommendation systems, and social psychology frameworks. This interdisciplinary study informs the understanding of MR's role in improving social presence, catalyzing novel social interactions, and enhancing the quality of interpersonal communication in the real world. We also discuss the challenges and barriers blocking the wide-spread adoption of social networking in MR, such as device constraints, privacy and accessibility concerns, and social norms. Through carefully structured, closed-environment experiments with diverse participants of varying levels of digital literacy, we measure the differences in social dynamics, frequency, quality, and duration of interactions, and levels of social anxiety between MR-enhanced, mobile-enhanced, and control condition participants.

## 1 INTRODUCTION

According to the U.S. Surgeon General's 2023 advisory, the most pressing, yet pre-existing, public health issue in 2023 was loneliness. The pervasive crisis of loneliness and isolation has severe implications for individual and societal health, increasing the risk of premature death by 26% and 29% respectively, akin to the mortality risk associated with smoking up to 15 cigarettes daily [25]. This report outlines the critical health consequences of insufficient social connection for a huge population of Americans, including heightened risks of heart disease, stroke, mental health disorders, and cognitive decline. Additionally, the impacts of loneliness and isolation were exacerbated by the COVID-19 pandemic [12], where only 39% of adults in the U.S. reported that they felt very connected to others [25].

The growing prevalence of social isolation and the modern innovations in mixed reality (MR) introduce a pressing need and timely opportunity for fostering more meaningful social connections. MR has potential impact across various industries, such as decision-making platform tools, entertainment, medicine, and education [1]. Additionally, MR applications have demonstrated promise in enhancing collaborative learning, facilitating remote work, and providing new platforms for entertainment [2, 45]. Advancements in the field can be observed through the various MR devices on the market, such as Microsoft HoloLens, Magic Leap, Meta Quest, Lenovo ThinkReality VRX, Varjo XR, Google Glass, and Apple Vision Pro, as well as the adoption of smart glasses, such as the Meta Ray-Bans [1, 2].

Despite the enthusiasm around mixed reality [34], the development of MR applications involves navigating complex design considerations, including spatial user interface design, multi-platform integration, and device-related technological limitations. The literature also points to several challenges for future research, including ethical concerns such as privacy, accessibility, and overcoming the digital divide.

In this paper, we introduce a hyper-local social platform to create in-person and long-lasting connections. This platform serves as a dynamic space for users to discover, connect, and engage with others nearby, where they can explore localized music content, join exclusive virtual spaces, discover unique collectibles, grow a local following, and build genuine connections within their communities. Our application aims to redefine social networking by focusing on hyper-local interactions to increase the quantity and quality of people's relationships. We leverage this platform in mixed reality to conduct an empirical study with human participants, to measure impacts on social dynamics and interaction. Outlined below are the major contributions that we expect the reader to learn from this research.

**Contributions:**
(1) Review of the SOTA for mixed reality technology and practices for addressing the primary blockers towards mass adoption.
(2) Empirical study with human participants to measure the impacts of mixed reality-enhancements on socialization.
(3) Hyper-local social media application for iOS and visionOS to catalyze novel interactions among colocated strangers.

## 2  BACKGROUND
Our work builds upon research in MR and social networking theories. We examine the intersection of mixed reality, social science, and recommendation systems, and discuss the challenges that are blocking mass MR adoption, in order to identify mechanisms that can help elevate social engagement.

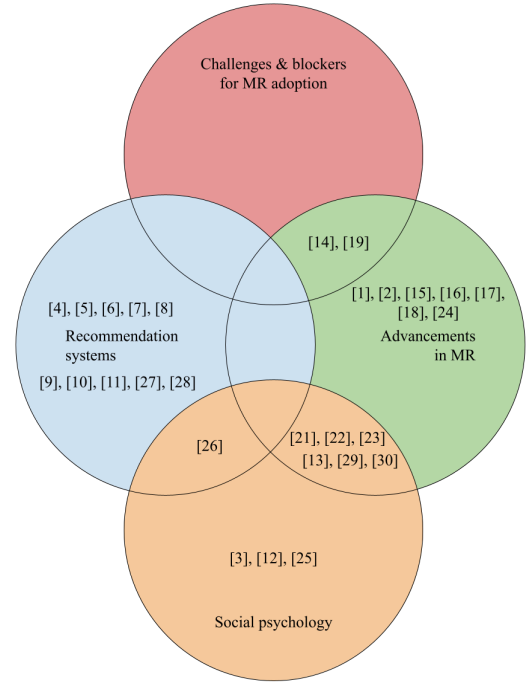

*Figure 1: Venn diagram on topic overlap*

**2.1 Advancements in Mixed Reality**
Mixed reality technology creates a hybrid environment where physical and virtual objects interact in real-time. This technology leverages advancements in augmented reality (AR) and virtual reality (VR) to enhance user experiences and interactions in various contexts, including social interaction. The evolution of MR has been propelled by advances in computer vision, graphical processing, display technologies, and input systems. This progression has enabled more natural and intuitive interactions within three-dimensional computer-generated environments, unlocking potential applications across various sectors including education, healthcare, and entertainment [1, 2, 43, 44, 45].

**3D Gaussian Splatting**
Gaussian splatting is a state-of-the-art technique in the field of computer graphics and data visualization that employs Gaussian functions to represent and blend data points in a spatial domain. Particularly in the context of point cloud visualization and volume rendering, this can be used to facilitate the creation of smooth, continuous visual representations from discrete samples. Central to its innovation is the method's ability to handle vast datasets efficiently, mitigating noise and enhancing visual clarity without compromising on computational performance.

Recent advancements have further refined Gaussian splatting by integrating machine learning algorithms and adaptive resolution techniques to optimize rendering based on the viewer's perspective and the density of the data points. These techniques can be used to not only improve the quality and speed of rendering but also expand the applications of Gaussian splatting across various domains, including medical imaging, scientific visualization, and AR systems.

**Near-Eye Displays**
Optical see-through lenses and near-eye displays (NEDs) are used extensively as the state of the art approach for enhancing mixed reality experiences. Significant progress has been made in improving resolution, expanding the field of view (FOV), and providing correct focus cues, enabling more immersive and realistic interactions with virtual content [14].

In the realm of display technologies, liquid crystal displays (LCDs) are used extensively across various applications, from television screens to mobile devices, due to their ability to produce high-resolution images with efficient power consumption [18]. However, the advent of NED technologies, particularly for mixed reality applications, has produced advancements beyond traditional, pure LCD capabilities.

Modern NEDs, including optical see-through systems, leverage cutting-edge liquid crystal optics to achieve compactness, high-resolution imagery, and dynamic focal adjustments in an ultra-compact form factor. Unlike LCDs, which are optimized for direct viewing on a flat panel, NEDs are designed to project images directly onto the user's retina, creating immersive visual experiences. This transition underscores a paradigm shift from traditional viewing platforms to personalized visual experiences, where requirements such as FOV, eye relief, and correct focus cues are handled more effectively. Recent innovations in liquid crystal technology for NEDs, such as polarization holography and phase modulation, have begun to address these requirements by offering NED devices that surpass the performance metrics of conventional LCDs in terms of immersion, portability, and user comfort.

As the technology progresses, the distinction between LCDs and NEDs continues to blur, with each advancement in liquid crystal optics paving the way for more sophisticated and immersive AR and VR experiences, such that traditional LCDs are not used as often.

**Waveguide Holography**
Innovations in waveguide optics, diffractive optical elements, and holographic techniques have led to more compact and efficient designs, reducing form factor while maintaining image quality and brightness [14]. This technology utilizes spatial light modulators for precise control of coherent light interactions to enable true 3D holographic projections. This method not only enhances resolution but also introduces a software-steerable large eyebox, significantly pushing the boundaries of immersive AR experiences [16].

**2.2 Social Psychology**
Theoretical frameworks within social psychology and computer-mediated communication offer valuable lenses through which to examine the impact of MR on social dynamics. For instance, MR can significantly enhance the sense of presence—a psychological state wherein virtual social interactions are perceived as authentic, potentially fostering deeper social connections. This notion aligns with the Social Presence Theory, introduced in the seminal paper by Biocca et al. in 2001, which posits that the medium through which communication occurs can affect the perceived intimacy and immediacy of the interaction, thereby influencing relational dynamics [3]. Studies have shown that MR can significantly affect social presence, the sense of being with another person in a virtual space, enhancing feelings of empathy, cooperation, and social connection [32].

In the context of MR, the social psychology field examines how social cues, presence, and interactions are perceived and influenced by immersive digital environments. This impact is critical in designing MR experiences that foster meaningful social interactions, bridging geographical and physical barriers [29, 30].

**Computer-mediated communication**
Studies have explored how digital platforms mediate social interactions and relationships. Computer-mediated communication (CMC) refers to any human communication that occurs through the use of two or more electronic devices. While traditionally focused on text-based communication, the field has expanded to include a broad range of multimedia elements and interfaces, with MR offering a new frontier for exploration. MR technologies extend traditional CMC paradigms by incorporating spatial and physical dimensions into digital communication, offering new avenues for research on social dynamics. The integration of MR into CMC studies focuses on understanding how augmented environments affect communication patterns, social norms, and user behavior.

## 2.4 Challenges & blockers
Despite the potential of MR to revolutionize social interaction, the remaining challenges include device constraints and ethical concerns. Future research must address these barriers to create inclusive, safe, and engaging MR social applications and experiences. Additionally, the integration of MR into social networks raises questions about user behavior, social norms, and non-verbal cues.

**Device constraints**
Despite the advancements outlined previously, challenges remain in achieving a perfect balance among all desired metrics without compromising device wearability or power efficiency. The pursuit of lighter, more power-efficient models with broader FOV and higher resolution continues to drive research and development, promising a future where AR NEDs could seamlessly integrate into daily life as replacements for traditional displays and become indispensable tools for professional, educational, and personal use.

The advancement in waveguide holography showcases a new and groundbreaking approach that combines waveguide displays with holographic technology to address the vergence-accommodation conflict and achieve a compact form factor in AR glasses. Additionally, adaptive focus systems for NEDs address the vergence-accommodation conflict by offering correct focus cues across varying distances, providing users with more comfortable and prolonged use. Eye tracking integration has further personalized the AR experience, optimizing the display based on the viewer's gaze and improving the eyebox to accommodate different users with minimal adjustments.

**Social norms**
MR CMC provides insight into remote interactions through non-verbal cues, such as gestures and spatial positioning, adding depth to remote interactions and presenting new challenges and opportunities for social connectivity. Social psychology provides insights into how individuals think, influence, and relate to one another within social interactions. MR technologies can impact these processes by altering perceptions of social cues, presence, and engagement. Our research in this area explores how MR affects group dynamics, social influence, and interpersonal relationships.

A recent paper published in February, 2024, "How Gaze Visualization Facilitates Initiation of Informal Communication in 3D Virtual Spaces" by Ichino et al. examines the role of non-verbal cues in catalyzing novel social interactions and models them as heat maps in 3D virtual environments [37]. This aligns with previous research by Paulos & Goodman, 2004 where they described the phenomenon of a familiar stranger as someone in a public place who we see and acknowledge non-verbally, but never physically interact with [35].

## 3  METHODOLOGY

To validate our hypotheses, we design a small-scale experiment focusing on a MR application that examines the traditional desktop and mobile philosophies of computer mediated communication for the mixed reality platform.

We employ a methodical framework predicated on an interdisciplinary approach, to investigate the nuanced impact of MR on in-person social interactions. We draw upon convergent paradigms of human-computer interaction, social psychology, computer hardware, communication studies, and ethics. This review synthesizes quantitative data with qualitative insights, offering a holistic examination of MR's social affordances and implications through user surveys and qualitative interviews.

Empirical studies have begun to explore the impact of extended reality on social interaction, with findings indicating potential benefits for education, healthcare, and workplace collaboration. Case studies of VR applications, such as social VR platforms and AR games [22], provide valuable insights into user engagement and community building. However, few studies have successfully explored native MR approaches to advance social connection.

### 3.1 Experiment design
This study systematically examines the impact of MR on social interaction. Drawing from seminal works such as Bailenson et al. [29] on social presence in virtual environments and Slater & Wilbur [30] on immersive virtual environments, this research aims to bridge the gap between MR technology's potential and its application in enhancing social connections.

**Participants**
A purposive sampling strategy will be utilized to recruit participants, ensuring a diverse representation across age, gender, and prior experience with MR technologies. Inspired by the methodology of Yee, Bailenson, Urbanek, Chang, & Merget [31], who investigated the psychological effects of avatar appearance in digital interactions, our participant selection criteria aim to encompass a broad spectrum of digital literacy levels.

**Approach**
The experiment will be structured as a shared learning environment designed to mimic real-world social interaction within a controlled setting where participants will be randomly assigned to one of three conditions:
1. MR-enhanced interaction condition: Participants will be equipped with head-mounted displays and motion-tracking technology.
2. Mobile interaction condition: Participants will be equipped with mobile devices as their medium of interaction.
3. Control condition: Participants will interact in a physical environment without any MR or mobile enhancements.

The social interaction task will be modeled after the experimental designs used by Fox, Arena, & Bailenson [32] who explored the influence of virtual embodiments on interpersonal trust and collaboration. Participants will be paired into groups of two, where they will receive bingo sheets with various prompts, ex. "has a pet", "recently solved a problem", "has felt lost", etc. These prompts will function as social triggers, which will be recorded throughout the session.

**Medium of communication**

Introducing the Hear application, which will function as the medium of communication for the MR-enhanced and mobile study participants. This application allows users to discover people nearby, message each other, and overcome initial social barriers associated with interacting with strangers nearby. This creates a closed channel of communication for the participants of this study.

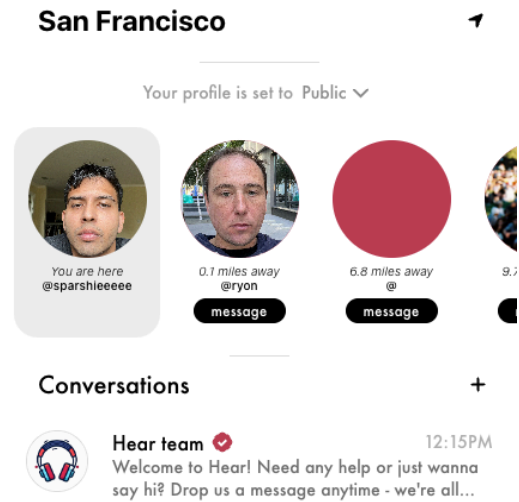

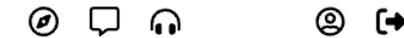

*Figure 3: Mobile interface of Hear*

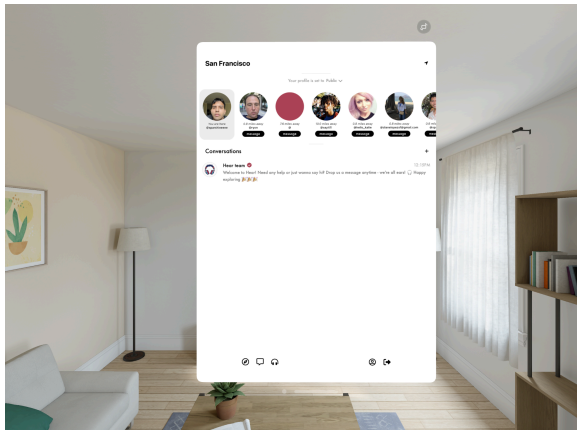

*Figure 2: Mixed Reality interface for Hear*

The mixed reality interface for Apple Vision Pro can be seen in Figure 2, which is utilized by the MR-enhanced participants, and the mobile interface seen in Figure 3 is utilized by the mobile participants.

Control participants were able to interact freely without restrictions, but mobile and MR-enhanced participants were required to interact initially through the Hear application, where they can see each other's profile pictures, username, bio, high-level interests, and hobbies.

**3.2 Data collection methods**

Quantitative data will be collected through pre- and post-experiment surveys measuring perceived social presence, co-presence, and task satisfaction, utilizing scales developed by Harms & Biocca [33].

Qualitative insights will be garnered through semi-structured interviews post-experiment,

focusing on participants' subjective experiences of social interaction within the MR environment. This approach is informed by the qualitative methodologies outlined by Quesnel & Riecke [34], emphasizing the phenomenological exploration of presence in immersive environments.

## 4 EVALUATION

There are four participants who contributed to this study, across a broad age range, equal gender distribution, and varying levels of digital literacy. Through the onboarding survey, we found that half of the participants described themselves as introverts and the other half as extroverts.

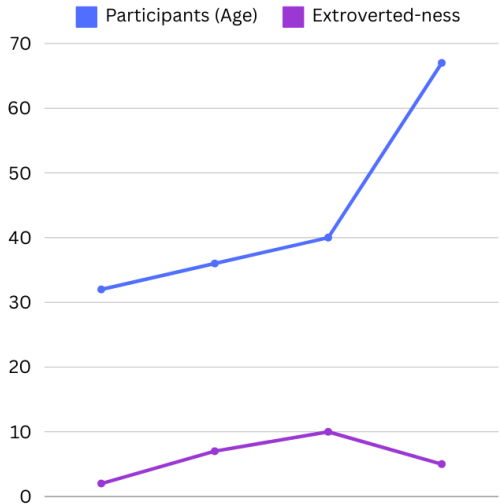

*Figure 4: Participants age vs. Self-reported extroverted-ness.*

The participants were asked to sign a liability waiver outlining responsibility for any damages to the mixed reality devices. The mixed reality devices used in this study included two Apple Vision Pro headsets, with the Hear application already installed.

### 4.1 Data analysis

The participants were paired into groups of two, in various combinations, such that each participant conversed with each of the other participants in five separate sessions, 3 minutes per session. The five configurations can be seen in Table 1, which depicts the interactions between participant 1 and participant 2, who represent real people in the study. The complete data outlines 30 sessions over six unique pairings.

The participants were given different bingo sheets to help nudge interaction during the sessions, with prompts of varying levels of closeness, from reasonably innocuous to overtly personal i.e. "has a hobby" vs. "has lost a family member or close personal friend". The participants were instructed to use the bingo sheets as a loose guide, given that their goal is simply to learn more about each other.

We measured the perceived quality of each of these sessions through the types of questions that the participants asked each other, the depth and frequency of interaction, and a qualitative post-experimental survey inquiring about the self-reported levels of comfort, anxiety, and connectedness. Frequency of interaction is defined as the number of times that the participant in that device group started a new topic of conversation, either from the bingo sheet or on their own. Conversation depth is measured by assigning numerical values to the depth of the questions on the bingo sheet, between 1 - 10 inclusive, that were asked by the participant in the device group. The average depth rating of each of the bingo sheets is five, where the deeper topics are balanced out by the more surface-level topics, such that the data only needs to be normalized by the number of interactions in the session, averaged over all of the sessions for the device group. Any other topics discussed outside of the bingo sheet are ranked for depth using the same discretionary criteria by the researchers.

The data in Table 2 represents the device group that the participant belonged to, regardless of which device group they were interacting with. This helps measure the ability of the participant to catalyze new social interactions depending on the medium of communication that they are using.

## 4.2 Observations

We use the metric of quality as a proxy for the level of social presence felt by the participants. Quality has a negative relationship with anxiety, so the average quality for a device group is calculated by taking the average of the following features - number of interactions, conversation depth, comfort, connectedness, and ten minus the average level of anxiety.

Since each of the participants were in each of the device groups for the same number of sessions, any biases generated by introversion/extroversion is eliminated. In this group, we did not observe that any single participant had a preference for or against any of the other participants based on personal similarity or experience. Due to the nature of the experiment, there are fewer sessions for the control device group, which may be allowing for outliers in the data to be over-represented.

The speed of response for the control group and the MR group was very similar, but some participants in the mobile group raised that they had a delay between their responses, despite being right next to each other, since they were the only group required to interface entirely through the device due to the nature of the experiment. This could impact the number of interactions, given that the session occurred over a limited time window of three minutes each.

## 5 RESULTS

Mediated-communication theories have been used to describe the differences between the print media, radio, and television. Computer-mediated communication (CMC) applies this theoretical framework to the internet, by the use of desktop computers, mobile phones, smart appliances, and now mixed reality devices. Social presence theory, which is often described as an extension and explanation of CMC, is defined as "the degree to which we as individuals perceive another as a real person and any interaction between the two of us as a relationship" [47]. In this research, we examine how social interactions mediated by mixed reality devices compare to conventional interactions, in-person or through mobile devices.

## 5.1 Social Presence

Through this experiment, we find the overall quality of social interactions for MR-enhanced participants is higher than mobile participants. Participants in the control group still show the highest quality of social interactions, though performing worse than the mobile group in specific categories, such as comfort and anxiety, and worse than the MR-enhanced group in other categories, such as conversation depth and connectedness. Using overall quality as a proxy for identifying perceived social presence, we can conclude that in-person social interaction observes the highest amount of social presence with a value of 5.7, MR-enhanced interactions perform second-best with a value of 5.0, and mobile interactions perform third with a value of 4.7, despite beating out the control group in specific categories.

## 5.2 Criteria for Peer Selection

After the experiment sessions, when given the opportunity to select the peer that they wanted to interact with, 100% of mixed-enhanced participants chose to interact with other participants wearing mixed reality devices or using mobile devices. This could indicate that the participants of this study preferred interacting with other people in the same condition as them - MR-enhanced, mobile-enhanced, or control. This underlying assumption was further confirmed by qualitative surveys conducted after the experiments, where all of the participants said they felt more comfortable talking to each other when both participants were wearing mixed reality devices rather than only one.

## 6 DISCUSSION

This study underscores the transformative potential of mixed reality in catalyzing social interactions. We found that frequency of interaction and social presence for the MR-enhanced participants was not drastically lower than the control group and that MR-enhanced participants were more likely to interact with other MR-enhanced participants. Additionally, we learned that social presence for MR-enhanced participants was higher than that of participants in the mobile group.

## AVAILABILITY

Our GitHub repository hear-web-mvp contains Swift, TypeScript, and Ruby scripts implementing the application used to mediate social interactions in mixed reality.

https://github.com/HearApp/web-mvp

## ACKNOWLEDGEMENTS

We thank Professor Gail Kaiser, PhD. for supporting the development of this project. Zain Ali provided assistance in conducting the research study with human participants, without whom the study could not have been completed. This work was supported in part by the School of Engineering and Applied Sciences at Columbia University in the City of New York. Any opinions, findings, and conclusions or recommendations expressed in this material are those of the authors and do not necessarily reflect the views of the supporting entities. We sincerely thank the participants of this study for making this work possible.

Catalyzing Social Interaction in Mixed Reality

Catalyzing Social Interaction in Mixed Reality**APPENDIX**

**Table 1:** *Experimental study between participant 1 & participant 2.*

| Session | Participant 1 | Participant 2 | Duration |
|---|---|---|---|
| 1 | MR-Enhanced | Mobile | 3 minutes |
| 2 | Control | Control | 3 minutes |
| 3 | MR-Enhanced | MR-Enhanced | 3 minutes |
| 4 | Mobile | MR-Enhanced | 3 minutes |
| 5 | Mobile | Mobile | 3 minutes |

**Table 2:** *Quality of social interaction for each device group.*

| Group | Number of interactions (avg) | Conversation depth (avg) | Comfort (avg) | Anxiety (avg) | Connectedness (avg) | Overall quality |
|---|---|---|---|---|---|---|
| Control | 5.5 | 3.7 | 6.5 | 4.3 | 7.0 | 5.7 |
| Mobile | 4.8 | 1.7 | 6.8 | 2.0 | 2.1 | 4.7 |
| MR | 5.2 | 5.4 | 3.3 | 6.5 | 7.8 | 5.0 |